\documentclass[%
reprint,
amsmath,amssymb,
aps,pra]{revtex4-1}

\usepackage{color}
\usepackage{graphicx}
\usepackage{dcolumn}
\usepackage{bm}
\usepackage{footmisc}
\usepackage{mathtools}
\usepackage{framed}
\usepackage{mathrsfs}
\usepackage{lipsum}

\newcommand{\C}{\mathbb{C}}

\def\d{{\rm d}}

\def\>{\rangle}
\def\<{\langle}

\newcommand{\st}[1]{\mathbf{#1}}
\newcommand{\bs}[1]{\boldsymbol{#1}}

\newcommand{\map}[1]{\mathcal{#1}}
\newcommand{\Tr}{\operatorname{Tr}}

\newcommand*{\YY}[1]{{\color{red} [YY: #1]}}
\newcommand*{\XZ}[1]{{\color{blue} [XZ: #1]}}

\begin{document}

\preprint{APS/123-QED}

\title{
Quantum Metrology with Indefinite Causal Order}

\author{Xiaobin Zhao$^{1,2}$, Yuxiang Yang$^{3}$ and Giulio Chiribella$^{1,2,4,5}$}\email{giulio@cs.hku.hk}
\affiliation{$^1$ Department of Computer Science, The University of Hong Kong,
Pok Fu Lam Road, Hong Kong 999077, China}
\affiliation{$^2$ The University of Hong Kong Shenzhen Institute of Research and Innovation, Yuexing 2nd Rd
Nanshan, Shenzhen 518057, China}
\affiliation{$^3$
Institute for Theoretical Physics, ETH Z\"urich, Z\"urich 8093, Switzerland}
\affiliation{$^4$ Department of Computer Science, University of Oxford, Parks Road, Oxford OX1 3QD,  United Kingdom}
\affiliation{$^5$ Perimeter Institute for Theoretical Physics, Caroline Street, Waterloo, Ontario N2L 2Y5, Canada}
	
\nopagebreak
	
\begin{abstract}
		
We address the study of quantum metrology enhanced by indefinite causal order, demonstrating a quadratic advantage in the estimation of the product of two average displacements in a continuous variable system. We prove that no setup where the displacements are used in a fixed order can have root-mean-square error vanishing faster than the Heisenberg limit $1/N$, where $N$ is the number of displacements contributing to the average. In stark contrast, we show that a setup that probes the displacements in a superposition of two alternative orders yields a root-mean-square error vanishing with super-Heisenberg scaling $1/N^2$, which we prove to be optimal among all superpositions of setups with definite causal order.   Our result opens up the study of new measurement setups where quantum processes are probed in an indefinite order, and suggests enhanced tests of the canonical commutation relations, with potential applications to quantum gravity.

\end{abstract}
\maketitle

The traditional formulation of quantum mechanics assumes that the order of physical processes is well defined. Recently, a number of works started exploring new scenarios where the causal order is indefinite
\cite{chiribella2009beyond,oreshkov2012quantum,colnaghi2012quantum,chiribella2013quantum,baumeler2014perfect,bisio2019theoretical}. 
This extension is motivated  by ideas in quantum gravity, where the order of  events   could be subject to quantum indefiniteness \cite{butterfield2001spacetime,hardy2007towards}, and has potential applications in quantum information,  where advantages have been found in channel discrimination tasks \cite{chiribella2012perfect,araujo2014computational}, non local games \cite{oreshkov2012quantum, baumeler2014perfect}, and communication complexity \cite{guerin2016exponential}.

A paradigmatic example of process with indefinite causal order is the quantum SWITCH \cite{chiribella2009beyond,chiribella2013quantum}, a higher-order operation that combines two input  gates in a quantum superposition of two alternative  orders.  When applied to two unitary gates $U_1$ and $U_2$, the quantum SWITCH generates  the controlled unitary gate
\begin{align}\label{QS}
S\left(U_1,U_2\right):=|0\>\<0|\otimes U_2U_1+|1\>\<1|\otimes U_1U_2  
\end{align}
by querying each of  the two   gates $\{U_1,U_2\}$ only once.  
Here first register on the right-hand side of Eq.\ (\ref{QS}) serves as a control of the  order. When  put in  a coherent superposition of the states $|0\>$ and $|1\>$, it induces a coherent superposition of the two alternative orders $U_1 U_2$ and $U_2 U_1$. 
The quantum SWITCH has been shown to offer a number of information-processing advantages \cite{chiribella2012perfect,araujo2014computational,guerin2016exponential} and has inspired experiments in quantum optics  \cite{procopio2015experimental,rubino2017experimental,goswami2018indefinite,guo2020experimental,wei2019experimental}, where the superposition of orders is reproduced by sending photons on a superposition of alternative paths \cite{chiribella2019quantum}.  Recently, it has stimulated an extension of Shannon theory to scenarios where the order of the communication channels is in a quantum superposition  \cite{ebler2018enhanced,salek2018quantum,chiribella2018indefinite}.  
 
In this work, we show that the quantum SWITCH can boost the precision of  quantum metrology, beating the limits associated with conventional schemes where processes are probed in a definite order. To illustrate this phenomenon,  we consider a situation where an experimenter has access to $2N$ black boxes, each acting on a harmonic oscillator, with the promise that the first $N$ boxes perform displacements generated by a given quadrature $ X$, and the second $N$ boxes perform displacements in the conjugate quadrature $ P$.
   Displacements performed by different boxes are independent, and   the  task is to measure the product of the average displacement in $ X$ and the average displacement in $ P$.  
   
   When the black boxes are used in a fixed order, we prove that the root mean square error (RMSE) cannot vanish faster than $f(E)/N$, where $f(E)$ is a function of the energy of the input states used to probe the black boxes.  The scaling $1/N$ is consistent with the Heisenberg limit  of quantum metrology \cite{giovannetti2006quantum}, applied to the estimation of the two average displacements in $X$ and $P$.  
     In stark contrast, we show that a setup using the quantum SWITCH can achieve an error vanishing with super-Heisenberg scaling  $1/N^2$, independently of the energy of the input states.     Our result demonstrates that a setup that probes a sequence of processes in a coherent superposition of alternative orders can extract more information than any setup where the order of the processes is fixed. Furthermore, we show that the scaling $1/N^2$, achieved by our concrete setup, is optimal among all setups obtained by superposing  causally ordered processes with bounded energy.

Our scenario can be described as follows. An experimenter has access to $2N$ black boxes, each implementing either a position displacement  $D_{x_j} = e^{-i x_j P }$ or a momentum displacement  $D_{p_k} = e^{i p_k X }$ ($j,k=1,\dots,N$), where $ X$ and $ P$ are the conjugate variables   $ X:=( a+ a^\dag)/\sqrt{2}$  and $ P:=i( a^\dag - a )/\sqrt{2}$, and  $ a$ and $ a^\dag$ satisfy the canonical commutation relation $[   a ,   a^\dag ] = {I}$.  
   The displacements $\{x_j\}$ and $\{p_k\}$ are unknown, and vary independently within the range $[x_{\min},  x_{\max}]$ and $[p_{\min},  p_{\max}]$, respectively.  
     The task is to estimate the product $A  :=  \overline x\cdot \overline p$      between  the average  displacements  $\overline x  :=  \sum_{j  =1}^N  x_j/N$ and   $\overline  p   :=  \sum_{j=1}^N  p_j/N$, by querying each black box only once in every run of the experiment.  For simplicity, we will assume that the average displacements $\overline x$ and $\overline  p$ are nonzero and converge to nonzero values  in the large $N$ limit.  

\begin{figure}[h]
\includegraphics[scale=0.31]{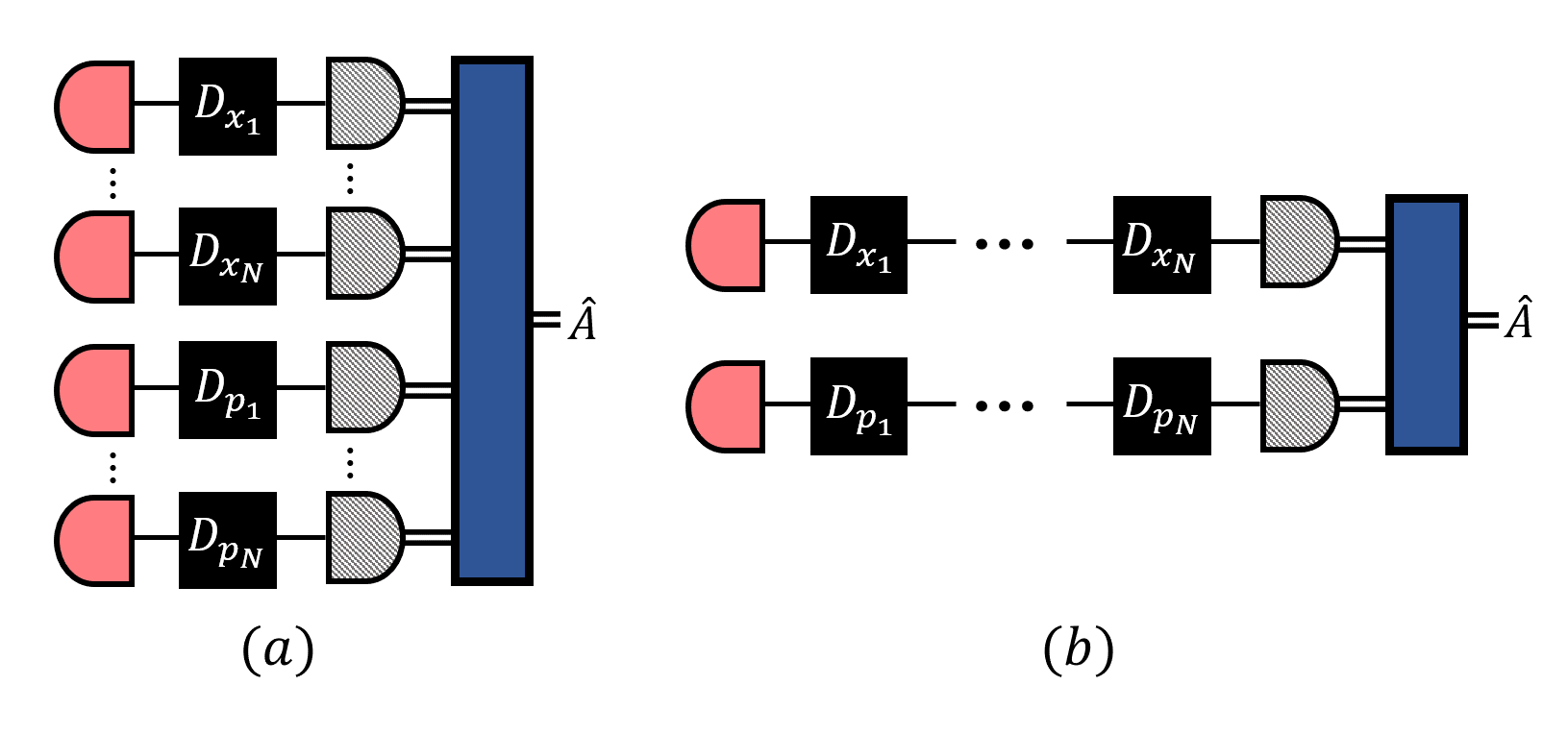}
\caption[]{ {\bf Two causally ordered schemes.}  
{\em (a) Parallel  scheme with  measurements of individual displacements.}     $2N$  independent probes, each with average energy bounded by $E$, are used to estimate 
 the $2N$ displacements $(x_i)_{i=1}^N$ and $(p_j)_{j=1}^N$.  The average displacements $\overline x  =  \sum_i x_i/N$ and $\overline p=  \sum_j  p_j/N$, and their product $A  =  \overline x \, \overline p$ are then computed by classical postprocessing. The RMSE of the scheme  has the standard quantum limit scaling $1/\sqrt  N$.   {\em (b) Sequential scheme with independent $x$ and $p$ measurements.}  The average displacements  $\overline x$ and $\overline p$ are measured directly by  applying  the total $x$ displacement $D_{x_1}  D_{x_2}  \cdots  D_{x_N}$ and the total $p$ displacement  $ D_{p_1} D_{p_2} \cdots D_{p_N}$ to two independent  probes, each with average energy bounded by  $E$.   The product $A  = \overline x \, \overline p$ is then computed by classical postprocessing.  The RMSE of this scheme has the Heisenberg scaling $1/N$. }
\label{fig:fixedordercircuit}
\end{figure}

The simplest way to estimate $A$ is to measure each displacement independently,  as illustrated in Fig. \ \ref{fig:fixedordercircuit}($a$). 
 A  bound on the RMSE follows immediately from the quantum Cram\'er-Rao bound \cite{helstrom1976quantum,holevo2011probabilistic,braunstein1994statistical}, which  can be applied to  the estimation of a displacement $z$,  yielding the 
    lower bound $\Delta z   \ge 1/  \sqrt{8\nu E}$, where  $E  :=   \<\psi  |   (X^2+ P^2)    |\psi\>/2 $ is the average energy of the probe state, and $\nu$ is the number of repetitions of the experiment ( see  Appendix \ref{APP:single_estimation} for a derivation.)   This bound implies that, once the energy $E$ has been fixed, the error in the estimation of a single displacement is a constant.    
    The error in the estimation of individual displacements then propagates to the estimation of the product, yielding an overall scaling   $1/\sqrt {\nu N}$,  corresponding to the standard quantum limit \cite{giovannetti2006quantum}.

   A better scaling  can be obtained if, instead of measuring each displacement separately, one directly measures the two average displacements $\overline x$ and $\overline p$,  by  applying  the total $x$ displacement $D_{N \overline x}  =  D_{x_1}  D_{x_2}  \cdots  D_{x_N}$ and the total $p$ displacement $D_{N \overline p}  =  D_{p_1} D_{p_2} \cdots D_{p_N}$ to two independent probes, each of  average energy $E$, as in Fig.  \ref{fig:fixedordercircuit}($b$).  In this case, the Cram\'er-Rao bound implies that the RMSE for each average displacement is lower bounded by $1/  (N\sqrt{8\nu  E})$, and therefore error propagation gives the RMSE scaling as $1/({\sqrt \nu} N)$ for the estimation of the product with any bounded energy $E$. 
   
   The $1/N$ scaling  corresponds to the Heisenberg limit for the estimation of the average displacements $\overline x$ and $\overline p$  \cite{giovannetti2006quantum}.  
   Later in the Letter we will prove that the scaling   $1/  N$ is optimal among all setups where the given  black boxes are probed in a definite order,  using a finite amount of energy.

\begin{figure}[h]
\includegraphics[scale=0.32]{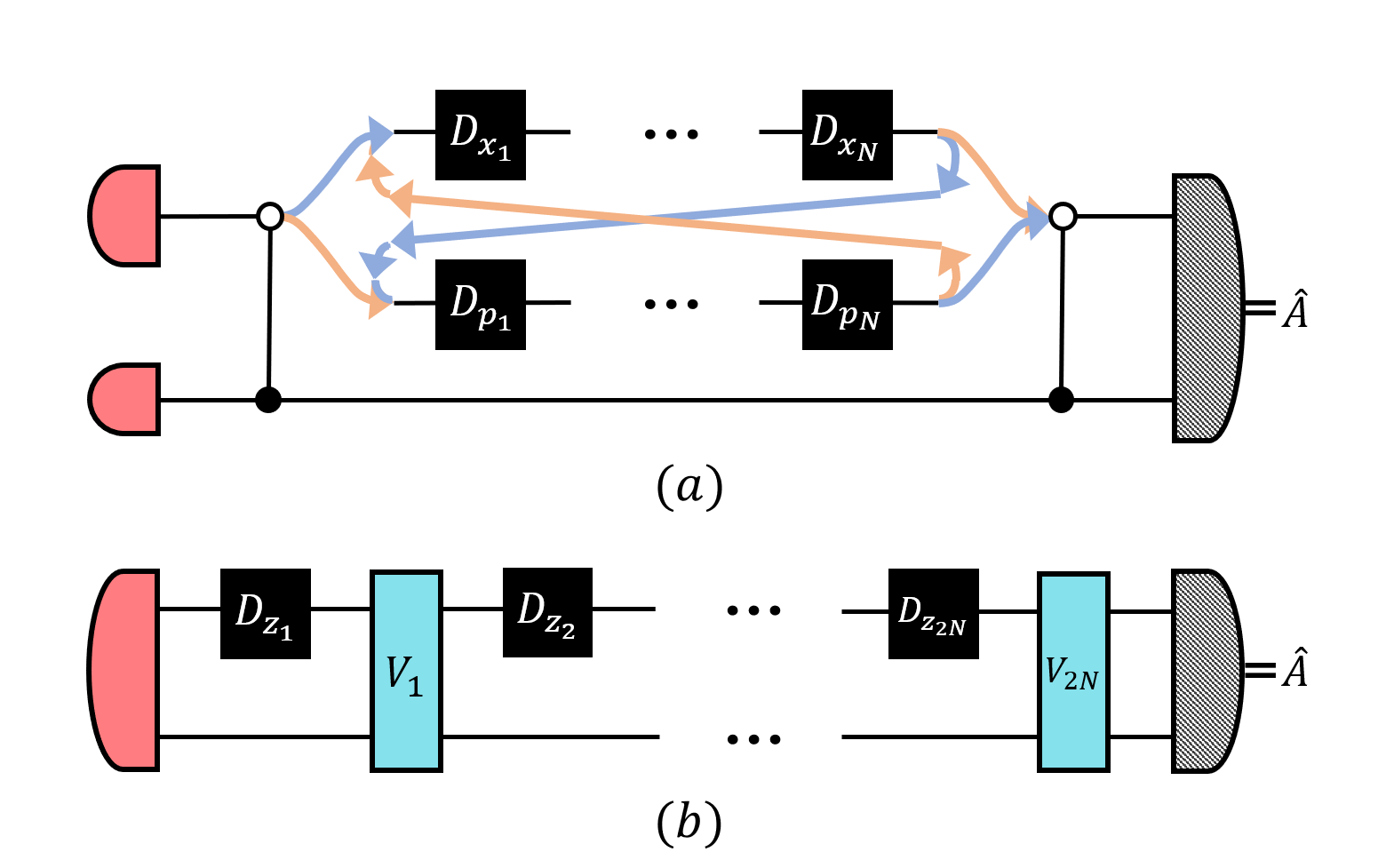}
\caption[]{ {\bf Definite {\em vs} indefinite order in  a quantum metrology setup.}  {\em  (a) Estimation scheme using the quantum SWITCH.}   The  total $x$ displacements $D_{x_1}  D_{x_2}  \cdots  D_{x_N}$ and  $p$ displacements   $D_{p_1} D_{p_2} \cdots D_{p_N}$ act in a coherent superposition of two alternative orders, controlled by the state of a control qubit.  If the  control is prepared in the state $|0\>$ ($|1\>$), the probe will experience the displacements in the order corresponding to the blue (orange) path. By preparing the probe in the minimum-energy state $|0\>$ and the control  qubit in the state $|+\>$,  this scheme achieves the super-Heisenberg scaling $1/N^2$ of the RMSE.   {\em (b) Generic causally ordered scheme.}  A probe and an auxiliary system are prepared in a generic state,  with average energy of the probe bounded by $E$.   Then, the probe undergoes a sequence of displacements,    arranged in  a fixed  order $(z_1,\dots,z_{2N})$, where  $(z_1,\dots,z_{2N})$ is an arbitrary permutation of the sequence $(x_1,\dots,x_N,p_1,\dots,p_N)$.  Each displacement operation   $z_i$ is followed by a  unitary gate  $V_i$, acting jointly on the probe and the auxiliary system.  Finally, a joint measurement is performed on the probe and the auxiliary system. Every estimation scheme of this form, including the schemes in Figs.  \ref{fig:fixedordercircuit}{\em (a)} and  \ref{fig:fixedordercircuit}{\em (b)}, must have the RMSE vanishing no faster than  $1/N$.  }
\label{fig:circuit34}
\end{figure}

We now show that a setup using the quantum SWITCH can achieve the super-Heisenberg scaling $1/N^2$.
  The setup creates a coherent superposition of two configurations: one where   all the $x$ displacements  are used first, and one where all the $p$ displacements are used first,  as in Fig.  \ref{fig:circuit34}{\em (a)}.  
 The process experienced by the probe is a unitary with a qubit control 
\begin{align}\label{QS2}
W&=|0\>\<0|\otimes \prod_{j=1}^{N} D_{p_j}  \prod_{j=1}^{N} D_{x_j}  +  |1\>\<1| \otimes \prod_{j=1}^{N} D_{x_j}  \prod_{j=1}^{N} D_{p_j} .
\end{align} 

Our scheme for estimating $A$ is  illustrated in Fig.  \ref{fig:circuit34}($a$). It  consists of the following steps:
\begin{enumerate}
\item[{(1)}] Prepare the control of the quantum SWITCH in the state $|+\>:=(|0\>+|1\>)/\sqrt{2}$.
\item[{(2)}] Prepare the probe in an arbitrary state $|\psi\>$, such as {\em e.g.} the minimum-energy state $|0\>$.
\item[{(3)}] Apply the gate  $W$  to the input state $|+\>\otimes |\psi\>$.
\item[{(4)}] Measure  the control using the projective measurement $\left\{|+\>\<+|,|-\>\<-|\right\}$ with $|-\>:=(|0\>-|1\>)/\sqrt{2}$.
\item[{(5)}] Repeat the above procedure for $\nu$ rounds and output the maximum likelihood estimate $\hat{A}:=\arg\max_A\log p(m_1,\dots,m_{\nu}|A)$, where  $m_j\in\{+,-\}$ is the $j$ th measurement outcome, and $p(m_1,\dots,m_{\nu}|A)$ is the probability of obtaining the measurement outcomes $\{m_1,\dots,m_\nu\}$ conditioned on the parameter being $A$.
\end{enumerate}

Using the Weyl relation $   e^{ip{X}} e^{-ix{P}}=e^{\text{i}xp{I}}e^{-ix{P}} e^{ip{X}}$,    
the output unitary of the SWITCH [Eq.\ (\ref{QS2})] can be cast into the  product form
\begin{align}
W&=  \left(|0\>\<0|+e^{iN^2A}|1\>\<1|\right)\otimes\left(\prod_{j=1}^{N} D_{p_j}  \prod_{j=1}^{N} D_{x_j}  \right).
\end{align}
Then, one can immediately see that the final state of the control qubit is $(|0\>+e^{iN^2A}|1\>)/\sqrt{2}$, and the probability of getting the outcome $\pm$ is  $p(\pm|A)=[1\pm\cos(N^2A)]/2$.

Since our estimator is unbiased, its RMSE  satisfies the Cram\'{e}r-Rao bound \cite{cramer1999mathematical,rao1992information,fisher1925theory}
\begin{align}\label{cramer}
\Delta A_{\rm switch}\ge\frac{1}{\sqrt{\nu F_A}}
\end{align}
where $F_A$ is the Fisher information of the parameter $A$, given by 
\begin{align}
F_A:=\sum_{m\in\{+,-\}}p(m|A)  \left[\frac{\partial \ln p(m|A)}{\partial A}\right]^2=N^4 \,.
\end{align} 
The Cram\'{e}r-Rao bound  [Eq. (\ref{cramer})] is achievable in the large $\nu$ limit, and we have the asymptotic equality
\begin{align}\label{mse-indef}
\Delta A_{\rm switch}  =\frac{1}{\sqrt{\nu}N^2} \, .
\end{align}
Hence,  the estimation scheme based on the quantum SWITCH  achieves the super-Heisenberg scaling $1/N^2$ in terms of the number of displacements contributing to the the average.    Notice that the $1/N^2$ scaling is independent  of the energy of the probe, meaning that the quantum SWITCH allows one to extract precise information even in the low-energy regime.

Our estimation scheme provides an accurate estimate for small values of  the parameter $   A$, {\em i.e.}, values  not exceeding the period of the functions $p(+ |A)$ and $p(-|A)$. Alternatively, our estimation scheme can be seen as a way to estimate the total phase $\phi:  =  \sum_{i,j}  x_i  p_j  \mod 2\pi $ with RMSE $\Delta  \phi_{\rm switch}= 1/\sqrt \nu $.  This scaling cannot be achieved with the causally ordered estimation scheme of Fig. \ref{fig:fixedordercircuit}{\em (b)}, because the total displacements  in $x$ and $p$  grow as $N$, and therefore error propagation  implies that the RMSE of their product grows as $N$, thus making the estimation of the phase $\phi$ unreliable whenever $N$ is large compared to $2\pi$.   More generally, we will see that no causally ordered scheme can achieve the RMSE scaling $\Delta  \phi  =  1/\sqrt {\nu}$.

 
 Note that our  scheme does not involve any measurement on the probe.  The scheme can be further improved by measuring the probe with a heterodyne measurement, whose measurement operators are projections on coherent states.  When the probe is initialized in a coherent state, such as the minimum-energy state $|0\>$, we show that our scheme can achieve  RMSE
\begin{align}\label{mse-indef-explicit}
\Delta A_{\rm switch}'& = \frac 1 {\sqrt \nu N^2}  \, \sqrt{\frac{{\overline x}^2+{\overline p}^2 }{    {\overline x}^2+{\overline p}^2+1/N^2 }} \, .
\end{align}
The  derivation of Eq.\ (\ref{mse-indef-explicit}) can be found in Appendix \ref{app:switch}.

We now show that the error scaling  $1/N^2$ cannot be achieved if the unknown displacements are used in a definite order.  Specifically, we will show that every estimation strategy with fixed order [see Fig. \ref{fig:circuit34} $(b)$] will have RMSE vanishing no faster than  $1/N$.  Suppose  that the first  displacement operation in the sequence is $D_{x_1}$.   In this case, every estimation scheme with fixed causal order  can also be used to estimate $A$ in the less challenging scenario where all the  displacements except  $x_1$ are known.   In this scenario, the RMSE  is simply $\Delta x_1/|\partial x_1/\partial A|=|\overline{p}|\Delta x_1/N$, where $\Delta x_1$ is the error in estimating $x_1$ from the displacement operation $D_{x_1}$.   Similarly, if the first displacement operation is $D_{p_1}$, one obtains  RMSE  $\Delta p_1/|\partial p_1/\partial A|=|\overline{x}|\Delta p_1/N$, where   $\Delta p_1$ is the error in estimating $p_1$ from the displacement operation $D_{p_1}$.    In general, the RMSE for the estimation of $A$ in  any fixed causal order is lower bounded as
\begin{align}\label{mse-fixed}
\Delta A_{\rm fixed}\ge\frac{\min_j\,|c_j|\cdot\Delta z_j}{N},
\end{align}
where $\{z_j\}$ are  the $2N$ displacements, and $c_j=\overline{p}$ ($\overline{x}$) if $z_j$ is a position (momentum) displacement.  
Since the RMSE in estimating a displacement $z_j$ is lower bounded by $1/\sqrt{8\nu E}$ 
 with $E$ being the initial  energy of the probe,  Eq.  (\ref{mse-fixed}) yields the bound 
\begin{align}\label{mse-fixed2} 
\Delta A_{\rm fixed}\ge\frac{\min\{|\overline{x}|,|\overline{p}|\}}{\sqrt{8\nu E}N}.
\end{align}
A more formal derivation of the bound  Eq. (\ref{mse-fixed2}) is provided in Appendix \ref{app:fixed}.

\begin{figure}[h]
\begin{center}
\includegraphics[scale=0.35]{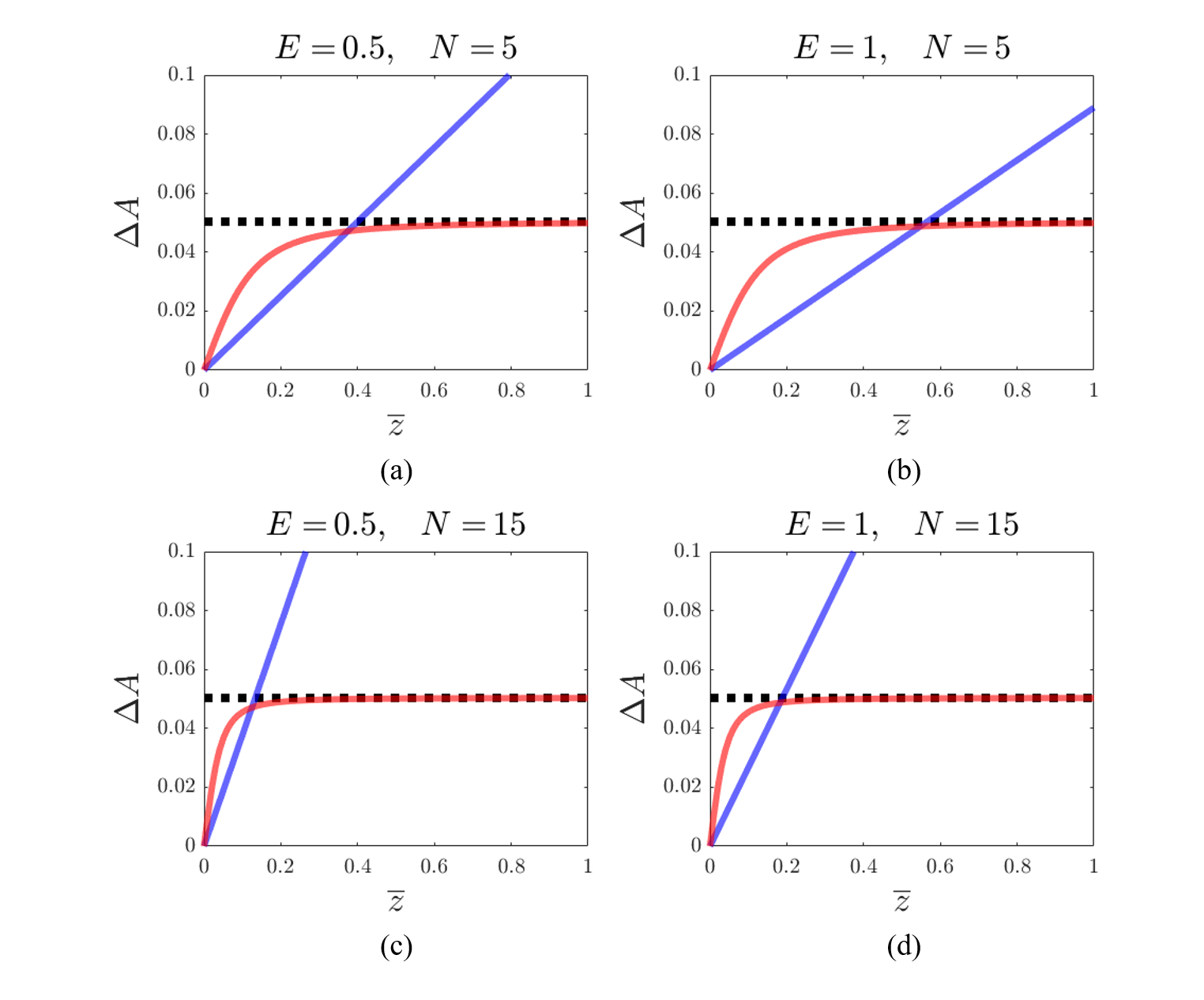}   
\end{center}
\caption[]{
{\bf   Definite {\em vs} indefinite order in the nonasymptotic regime.}   The RMSE achievable with the quantum SWITCH  is plotted against the lower bound to the RMSE for every estimation scheme with definite causal order. The four plots correspond to the parameter values   $|\overline x|=|\overline p|=\overline z>0$, $\nu=10$, and (a) $E =0.5, N=5$; (b) $E =1, N=5$; (c)  $E =0.5, N=15$; (d) $E =1, N=15$. The $y$ axis shows the RMSE $\Delta A$ in units of $2\pi/N^2$.    The solid red lines show  the  RMSE $\Delta A_{\rm switch}'$, achievable by measuring the probe and the control [Eq.~(\ref{mse-indef-explicit})].  The dashed lines show the RMSE $\Delta A_{\rm switch}$, achievable by measuring the control alone [Eq.~(\ref{mse-indef})]. The blue lines show  the lower bound of the RMSE $\Delta A_{\rm fixed}$ [Eq.~(\ref{mse-fixed2})].}
\label{fig:difference}
\end{figure}

The advantage of indefinite causal order  can immediately be identified when comparing the RMSEs Eqs.   (\ref{mse-indef-explicit})   and  (\ref{mse-fixed2}). Using a quantum SWITCH, the error  vanishes as $1/N^2$ instead of $1/N$. 
In terms of the phase $\phi   =   N^2 A \mod 2\pi$,  the quantum SWITCH offers RMSE scaling as $1/\sqrt  \nu$ with the number of repetitions of the experiment, while every scheme with definite causal order  has RMSE scaling at best as $N/\sqrt \nu$ in the $\nu \gg N$ regime.  In Fig. \ref{fig:difference} we  compare   the RMSE Eq.  (\ref{mse-indef-explicit})  with the lower bound Eq.  (\ref{mse-fixed2})  for various  values of $N$ and $E$.

 A natural question is whether more general forms of indefinite causal order, other than the quantum SWITCH, can beat the scaling $1/N^2$.  As it turns out, the answer is negative  for  all superpositions of definite causal orders. 
  The argument can be sketched as follows.  The RMSE in the estimation of $A$   is lower bounded by the RMSE in the situation where all displacements except one (say $x_1$) are known.   In that case, we have $\Delta A = \overline p   \,  \Delta x_1/N$.   We then show that no superposition of causal orders with bounded energy can achieve RMSE $\Delta x_1$  vanishing faster  than   $1/N$.  Putting everything together, this means that the RMSE for the estimation of $A$ cannot vanish faster than $N^2$ (see Appendix \ref{APP:one_use_energy} for the full argument.) 

Our  protocol suggests a way to test modifications  of the canonical commutation relations, such as those envisaged in certain theories of quantum gravity \cite{garay1995quantum,szabo2003quantum,pikovski2012probing,kempf1995hilbert}. 
For example, Ref.\ \cite{kempf1995hilbert} argues that the commutation relation should be replaced by $ [ X, P]   =i     \,  \left(     I + \beta  \,   P^2\right)$, where $\beta\ll 1$ is a suitable coefficient.    Using the quantum SWITCH setup one can in principle create the superposition  
\begin{align}
|\Psi\>    =   \frac{ \left(  I\otimes D_p   D_x  \right)  \,  ( |0\>  \otimes |\psi\> +    |1\>\otimes  U |\psi\> )   }{\sqrt 2}  
\end{align}
where $U$ is the unitary operator  
\begin{align}
\nonumber U  &=     D_{-x}  D_{-p}    D_x  D_p \\
&=   e^{  - i  xp   }e^{-i \beta x\left( p P^2+p^2  P +\frac 1 3 p^3 \right)}    +   O(  \beta^{  2})\, .
\end{align}
Choosing the state $|\psi\>$ to be close to an eigenstate of the momentum operator, we then obtain the state $|\Psi\>   \approx  D_x   D_p    |\psi\>  \otimes   (|0\>  +  e^{-i xp[1+(7/3)\beta p^2]}  |1\>  )/\sqrt 2$.  If the size of the displacements grows linearly, namely $x  =  N  \overline x$ and $p =  N \overline p$ for two fixed values $\overline x$ and $\overline p$, then the  constant $\beta$ can be measured with  RMSE  scaling as  $1/N^{4}$.  In other words, our scheme offers a favorable  scaling with the size of the displacements.

Other theories of quantum gravity \cite{szabo2003quantum} exhibit noncommutativity of the position operators associated with different   Cartesian coordinates. For example, the position operators $ X $ and $ Y$ can  become  conjugate variables, satisfying the canonical commutation relation $[  X,  Y ]  =  i c_{xy} I$ where $c_{xy}$ is a small constant.  Therefore, in this scenario protocol could  in principle offer  a way to measure the constant and to discover small amounts of noncommutativity of the two coordinates $ X$ and $ Y$.

These potential applications motivate the search for experimental implementations of our setup. For discrete variables, the quantum SWITCH can be reproduced on photonic systems using superpositions of paths \cite{procopio2015experimental,rubino2017experimental,guo2020experimental}. For continuous variables, Ref.  \cite{giacomini2016indefinite} suggests that a quantum SWITCH could be implemented in new experiments with  Gaussian quantum optics.  However,   no  photonic  realization of  the continuous-variable quantum SWITCH has been proposed to date. Alternatively, we suggest that the continuous-variable quantum SWITCH could be implemented with massive  particles with a continuous-variable internal degree of freedom, using the path of the particle to control the order of different displacement operations.  
For example, the internal degrees of freedom could be  the vibrational modes of a molecule or  the internal states  of a Bose-Einstein condensate. Another alternative is to reproduce our setup in  ion trap systems, where the spin and the axial mode of motion of an ion can be coupled together in a way that implements the control-unitary gates $U_j= |0\>\<0| \otimes  D_{x_j}    + |1\>\<1| \otimes  D_{x_j}^\dag $  and   $V_j  =  |0\>\<0|   \otimes D_{x_j}^\dag  +  |1\>\<1|  \otimes  D_{x_j}  $ \cite{sorensen1999quantum,haljan2005spin}. In this scenario, the quantum SWITCH can be simulated by first applying all the gates $U_j$ (with $j$ running from $1$ to $n$), then all the displacements $D_{p_j}$, and finally  all the gates $V_j$. Overall, this sequence of gates results in the  gate $(|0\>\<0|  +  e^{2N^2A i} |1\>\<1|)  \otimes  D_{N \overline x}^\dag   D_{N\overline p}  D_{  N  \overline x}$, from which the parameter $A$ can be estimated with RMSE $\Delta A   =   1/(2\sqrt{\nu}N^2)$.

In summary, we showed the quantum metrology schemes using indefinite causal orders can sometime outperform the standard schemes where quantum processes are probed in a definite order.  Specifically, we showed that every estimation scheme that probes $N$ pairs of displacements in a definite order has an error vanishing no faster than  $1/N$  for the estimation of the product of the average displacements. Instead, we showed that an estimation scheme using the quantum SWITCH achieves the enhanced scaling $1/N^2$.   Our result opens up a new area of research on the study of quantum metrology schemes powered by indefinite causal order.  
 
\medskip

This work is supported by the by National Science
Foundation of China through Grant No. 11675136, by Hong Kong Research Grant Council
through Grants No. 17326616 and No. 17300918, by the Croucher Foundation, and by the Swiss National Science Foundation via the National Center for Competence in Research ``QSIT" as well as Project No.\ 200020\_165843, and by the ETH Pauli Center for Theoretical Studies. This publication was made possible through the support of the ID 61466 Grant from the John Templeton Foundation, as part of the “The Quantum Information Structure of Spacetime (QISS)” Project (qiss.fr). The opinions expressed in this publication are those of the author(s) and do not necessarily reflect the views of the John Templeton Foundation.” 
 We thank M Hayashi for helpful comments on an earlier version of this manuscript.   
		
{\em Note added:} Recently, we found two  studies on the application of the quantum SWITCH  in quantum thermometry \cite{mukhopadhyay2018superposition} and channel identification \cite{frey2019indefinite}. These works showed  an increase of the quantum Fisher information by a constant amount  when the order of two channels is put in a coherent superposition, but did not address the comparison with the performances of arbitrary schemes with definite causal order. 

\bibliographystyle{apsrev4-1}
\bibliography{reference}

\begin{widetext}

\appendix

\section{Error bound on estimating a single displacement}\label{APP:single_estimation}

Here we derive a lower bound on the RMSE for the estimation of a displacement $z$, generated  by a generic  quadrature $X_\varphi  = (e^{i\varphi}  \,  a  + e^{-i\varphi} \,  a^\dag)/\sqrt 2$. For the applications in the main text,  $z=  x$  and $\varphi  =  -\pi/2$, or $z=p$ and $\varphi  = \pi $. 

We denote the displacement operator  by $D_z  =  \exp[- i  z X_\varphi]$.  Consider any unbiased quantum estimator  that estimates the parameter $z$ from the state $|\psi_z\>=D_z|\psi \>$. 
By the quantum Cram\'er-Rao bound \cite{helstrom1976quantum,holevo2011probabilistic,braunstein1994statistical}, the RMSE of can be lower bounded as
\[\Delta z   \ge \frac{1}{  \sqrt{\nu F_z}},\]
where $\nu$ is the number of repetitions of the experiment, and $F_z:=4\<\psi|X_{\varphi}^2|\psi\>-4\<\psi|  X_{\varphi}|\psi\>^2$ is the (symmetric logarithmic derivative) Fisher information of the parameter $z$. 
 
Furthermore, one has the bound  $F_z\le 4\<\psi|X_{\varphi}^2|\psi\> \le 4\<\psi|\left(X_{\varphi}^2+P_{\varphi}^2\right)|\psi\>$, with $P_\varphi  :  =  X_{\varphi+ \frac {\pi} 2}$.  Note that the sum $ X_{\varphi}^2+P_{\varphi}^2$ is independent of $\varphi$, and is equal to twice the Hamiltonian of the harmonic oscillator.  
All together, the above  bounds give 
\begin{align*}\Delta z &  \ge \frac{1}{ \sqrt{ 4\nu \<\psi|\left(X^2+P^2\right)|\psi\>}}\\
&=\frac 1{\sqrt{8\nu E }},
\end{align*}
where $E:=\frac 1 2 \<\psi|(X^2+P^2)|\psi\>$ is the average energy of the probe state $|\psi\>$.

\section{Proof of Eq. (\ref{mse-indef-explicit})}\label{app:switch}

The quantum SWITCH generates the unitary gate $W  =   |0\>\<0|  \otimes D_{N  \overline p}  D_{N\overline x}  +  |1\>\<1|  \otimes D_{N  \overline x}  D_{N\overline p}$. 
When the  state  $|+\>\otimes |\psi\>$ is input to this gate,  the output state is
\begin{align}
|\psi^{\rm switch}_A\>=\frac{|0\>+e^{iN^2A}|1\>}{\sqrt 2}  \otimes  D_{N\overline p}  D_{N\overline x}|\psi\>.
\end{align}
Suppose that the state $|\psi\>$ is  the minimum-energy state $|0\>$ and that, after the gate $W$ has acted, the control is measured in the basis $\{  |+\>  ,  |-\>\}$, while  the probe undergoes a measurement with   operators  $P_\beta:  = \frac 1 \pi   |\beta\>\<\beta|   $, where $|\beta\>$ is the  coherent state $|\beta\>:=e^{\beta a^\dag -\beta^* a}|0\> \, ,\beta \in \C$ and the measurement satisfies the normalization condition $\int  \, \d^2 \beta \,   P(\beta) =   I$.  The joint probability distribution of the   outcomes $ (\pm,\beta)$ has density
\begin{align}
p(\pm,\beta|A)&\nonumber =\<\psi^{\rm switch}_A| \left( |\pm\>\<\pm|\otimes P_\beta  \right)|\psi^{\rm switch}_A\>\\
&=\frac 1 {2\pi}  \,  \big[  1\pm\cos(N^2A)\big] \,  e^{-\left|\frac{N}{\sqrt 2}(\overline x+i\overline p)-\beta \right|^2}. 
\end{align}

The outcomes  $ (\pm,\beta)$  are then used to estimate the parameter $A$, using the maximum likelihood estimator.    The precision of the estimate is constrained by the Cram\'er-Rao bound, expressed in terms of the Fisher information matrix  $F$, computed by  taking  $A$ and $\overline x$ as independent parameters (when this is done, the parameter $\overline p$ can be expressed as $\overline p  =  A/\overline x$).    
 The entries of the Fisher information matrix are 
\begin{align}
F_{j,k} &=\sum_{\pm}\int d^2 \beta \frac { \partial_j\, p(\pm,\beta|A)  \,   \partial_k\, p(\pm,\beta|A)} {p (\pm,\beta| A)},
\end{align}
where  $j$ and $k$ take  values 1 or 2, and the partial derivatives are defined as $\partial_1:=\partial/\partial A, \partial_2:=\partial/\partial \overline x $. 
Explicitly, the derivatives can be written as follows
\begin{align}
\partial_1\, p(\pm,\beta|A)&=p(\pm,\beta|A)\left\{\frac{\mp N^2 \sin\left(N^2 A\right)}{1\pm\cos\left(N^2A\right)}+\frac{iN}{\sqrt 2\overline x} \left(i\sqrt 2N\overline p+\beta^* -\beta \right)\right\}\\
\partial_2\, p(\pm,\beta|A)&=-p(\pm,\beta|A)\frac{N}{\sqrt 2}\left\{\sqrt 2N\overline x-\sqrt 2\frac{\overline p^2}{\overline x}N-\left(1-i\frac{A}{\overline x^2}\right)\beta^*-\left(1+i\frac{A}{\overline x^2}\right)\beta\right\} \, ,
\end{align}
and the Fisher information matrix has the expression
\begin{align}
F&=\left[\begin{matrix}
N^4+\frac{N^2}{\overline x^2}&-\frac{N^2\overline p}{\overline x^2}\\
-\frac{N^2\overline p}{\overline x^2}&N^2 +\frac{N^2\overline p^2 }{\overline x^2 }
\end{matrix}
\right] .
\end{align}

Now, the Cram\'er-Rao bound   reads
\begin{align}\label{CRAO}
\Delta A  \ge  \sqrt{ \frac{ (F^{-1})_{1,1}}\nu}   = \frac{1}{\sqrt \nu  \,   N^2} \,  \sqrt{\frac {  \overline x^2  +  \overline p^2}  {   \overline x^2  +  \overline p^2+ 1/N^2}} \, ,
\end{align}
where $\nu$ is the total number of repetitions of the experiment.  In the asymptotic limit $\nu  \to \infty$, the MLE is known to achieve the bound (see, for instance, Ref.\ \cite[Page 63]{van2000asymptotic}).    Since we adopted the MLE in our estimation strategy, this proves Equation (7) in the main text.

\section{Proof of Eq. (\ref{mse-fixed2})}\label{app:fixed}

For an estimation scheme with fixed order, we denote by $\vec{z}:=(z_1,\dots,z_{2N})$ the  permutation of the vector $(x_1,\dots,x_N,p_1,\dots,p_N)$ corresponding to  the order in which the displacements  are queried.
The most general estimation scheme with fixed order is specified by a quantum circuit,  consisting of the preparation of  the probe and an auxiliary system in a  joint input state $|\psi\>$, followed by the execution of  the unknown displacements $D_{z_1}, \, D_{z_2} \, , \dots,  D_{z_{2N}}$, interspersed by a sequence of fixed  unitary gates $V_1,V_2, \dots,  V_{2N}$, as illustrated in Fig. 2(b) of the main text.   The overall  output state is 
\begin{align}\label{state}
|\psi_{\vec{z}} \>:=  \prod_{l=1}^{2N} V_l (D_{z_l}\otimes I_{\rm aux} ) | \psi \> \, , 
\end{align}
where $I_{\rm aux}$ is the identity on the auxiliary system.  

Now, suppose that all the displacements except $z_1$ are fixed and known to the experimenter. Then, the vector $\vec z$ can be replaced by the vector $\vec{z}_A:=(A,z_2,\dots,z_{2N})$  in the parametrization of the output state, which can be denoted as $|\psi_{\vec{z}_A}\>  =  U_{\vec{z}_A}  \,  |\psi\>$ for a suitable unitary gate $U_{\vec{z}_A}$. 
In this parametrization, the quantum Cram\'er-Rao bound \cite{helstrom1976quantum,holevo2011probabilistic,braunstein1994statistical} implies that the RMSE of any unbiased estimator is lower bounded by
\begin{align}\label{cramerrao}
\Delta A_{\rm fixed}\ge \frac{1}{\sqrt{\nu F^{\rm Q}_A}}
\end{align}
where  $\nu$ is the number of repetitions of the experiment and  $F^{\rm Q}_A$ is the (symmetric logarithmic derivative) quantum Fisher information, given by 
\begin{align}\label{FA}
F^{\rm Q}_A=4\left(\<\psi|{G}_A^2|\psi\>-\<\psi|{G}_A|\psi\>^2\right),
\end{align}
where ${G}_A:=i(\partial U_{\vec{z}_A}/\partial A)U_{\vec{z}_A}^\dag$ is the generator of $U_{\vec{z}_A}$. Explicitly, we have
\begin{align}
{G}_A=\frac{N}{c_1}\cdot{G}_{z_1}
\end{align}
where ${G}_{z_1}:=i(\partial D_{z_1}/\partial z_1)D_{z_1}^\dag$ denotes the generator of $z_1$, which is either ${X}\otimes I_{\rm aux}$ or ${P}\otimes I_{\rm aux}$, depending on which type of displacement $z_1$ is,  and $c_1$ is either $\overline{x}$ or $\overline{p}$, depending on which type of displacement $z_1$ is.  

Substituting into Eq.\ (\ref{FA}), the quantum Fisher information for the parameter $A$ is then
\begin{align}\label{QFI_relation}
F^{\rm Q}_A=\left(\frac{N}{c_1}\right)^2F^{\rm Q}_{z_1}\qquad F^{\rm Q}_{z_1}:=4\left(\<\psi|{G}_{z_1}^2|\psi\>-\<\psi|{G}_{z_1}|\psi\>^2\right)
\end{align}
where $F^{\rm Q}_{z_1}$ is the quantum Fisher information of the state $D_{z_1}|\psi\>$ with respect to the parameter $z_1$.  Since $G_{z_1}$ is equal to either   $X\otimes I_{\rm aux}$ or $P\otimes I_{\rm aux}$,  one has the bound 
\begin{align}\label{simplebound}
\nonumber F^{\rm Q}_{z_1} &  \le 4\,   \<\psi|{G}_{z_1}^2|\psi\>  \\
\nonumber  & \le      4\,   \<\psi|   \,  (X^2 +  P^2) \otimes I_{\rm aux}  \,  |\psi\>   \\
  &  =8 \,  E \, ,
  \end{align}
where $E=  \Tr  \left[  \frac{X^2 + P^2}2   \,\rho\right]$ is the average energy of the  probe state $\rho:  =  \Tr_{\rm aux}  [  |\psi\>\<\psi|]$, $\Tr_{\rm aux}$ denoting the partial trace over the auxiliary system.

    Inserting Equations (\ref{QFI_relation}) and (\ref{simplebound})  into the quantum Cram\'{e}r-Rao bound, we get  the bound 
\begin{align}
\nonumber \Delta A_{\rm fixed}  &\ge\frac{|c_1|}{N}\cdot\frac{1}{\sqrt{\nu F^{\rm Q}_{z_1}}}  \\
 \nonumber  &  \ge  \frac{|c_1|}{N}\cdot\frac{1}{\sqrt {8\nu E}}\\
 &\ge  \frac{   \min\{|\overline x|,|\overline p|\}}{  N  \sqrt{ 8 \nu E}}     \, . \label{rmse-fixed}
\end{align}
The above bound proves Equation  (\ref{mse-fixed2}) of the main text.   We remark that the bound (\ref{rmse-fixed}) is not tight, since in general the fact that the   displacements $z_2,\dots,z_{2N}$ are unknown increases the error in the estimation of $A$. Nevertheless, the scaling $1/(N\sqrt{\nu E})$ is achievable by separately estimating the average displacements $\overline x$ and $\overline p$, as discussed in the main text.

\section{Lower bound on the RMSE for arbitrary superpositions of causal orders}\label{APP:one_use_energy}

 Here we show that no superposition of setups with definite causal order can achieve a better RMSE scaling (with respect to $N$) than the scheme shown in the main text.  For this purpose, we derive a lower bound on the RMSE  for the easier task of estimating a single  displacement, say  $x_1$,  while  all the other displacements are known. In this scenario, the scheme for estimating the product $A  =  \overline x\, \overline p$ induces  a scheme for estimating $x_1$. In turn, such effective scheme can be decomposed into the preparation of a probe state, the application of the unknown gate $D(x_1)$, and a measurement on the output.  Our strategy will be to derive a limit on the scaling of  energy in the probe state.   Since the RMSE for estimating a displacement is related to the energy of the probe  [see Eq.\ (\ref{rmse-fixed})], the bound on the energy will lead directly to  a bound on the RMSE.

\subsection{Energy increase due to a single displacement}
We now start by  analyzing how the energy  increases under the action of  a single displacement operation. The displacement can be either a position displacement or a momentum displacement, and the amount of displacement is assumed to be either in the range $[x_{\min},  x_{\max}]$ or in the range $[p_{\min}, p_{\max}]$, as in the main text.

Consider an arbitrary  state $|\psi\>$ of the probe  and an auxiliary system, with a generic Hamiltonian $H_{\rm aux}$. Without loss of generality, we assume that the ground state of $H_{\rm aux}$ has zero energy, so that all the other eigenstates have positive energies.    The total energy of the system is  
\begin{align}
E_{\rm in}:&=     \<\psi|    H_{\rm tot}  |\psi\> \, ,   \qquad   H_{\rm tot}:  =  \frac {X^2+P^2}2 +  H_{\rm aux} \, ,
\end{align}
where we are omitting identity operators on the systems where the operators do not act (for example, we write $X^2$ for $X^2 \otimes  I_{\rm aux}$).

Suppose that the displacement is a position displacement  $D_x$.    After its  action,  the total energy becomes 
\begin{align}
\nonumber
 E_{\rm out}
&= \<\psi|   D_x^\dag \,     H_{\rm tot} \, D_x  |\psi\>\\
\nonumber
 &= \<\psi|    \,  \left[  \frac{ (X  + x)^2+P^2} 2 +  H_{\rm aux} \right]\,  |\psi\>\\
&=  E_{\rm in }+\frac{x^2}2+ x \, \<\psi| X |\psi\>\,.
\end{align}
Using the relation $\<\psi  |  X  |\psi\>   \le   \sqrt{  \<\psi  |  X^2 |\psi\>} \le   \sqrt{  \<\psi  |  (X^2  +  P^2) |\psi\>}  $ we obtain  the bound
\begin{align}
\nonumber
 E_{\rm out} & \le    E_{\rm in}   +  \frac{x^2}2   +   x  \sqrt{  2(  E_{\rm in}   -  \<\psi  |  H_{\rm aux}  |\psi\>  )}   \\
\nonumber &\le   E_{\rm in}   +  \frac{x^2}2   +   x  \sqrt{  2   E_{\rm in}   } \\
\nonumber 
& =      \left( \sqrt {E_{\rm in}}   +  \frac x {\sqrt 2}  \right)^2 \\
\label{energyb} & \le       \left( \sqrt {E_{\rm in}}   +  \frac {z_{\max}} {\sqrt 2}  \right)^2   \, , 
\end{align}
with $z_{\max}:=\max\{|x_{\min}|,|x_{\max}|,|p_{\min}|,|p_{\max}|\}$.   If the displacement is a momentum displacement, the  bound (\ref{energyb}) still holds.


\subsection{Energy increase in a causally ordered scheme with bounded energy}  

Consider a causally ordered scheme where the probe is prepared in a given input state $|\psi\>$, possibly entangled with an auxiliary system, and undergoes $2N$ displacements  $\st z =( z_1,\dots, z_{2N})$, interspersed with the fixed gates $\{  V_{j}\}_{j=1}^{2N}$.    We assume that the estimation scheme requires a bounded amount of energy, which implies that the energy of the input state $|\psi\>$ is bounded, and that each of the intermediate gates $\{  V_{j}\}_{j=1}^{2N}$ can be implemented with a bounded amount of energy.    We stress that, if no bound on the energy is assumed, then the displacements $( z_1,\dots, z_{2N})$ can already be estimated with arbitrarily precision by the naive causally-ordered circuit each displacement is measured individually.  

The energy requirement of a quantum gate is the amount of energy that should be supplied by a battery in order to implement the gate using only energy-preserving operations.   The energy requirement is lower bounded by the maximum amount of energy increase that the gate can induce on a generic input state  \cite{chiribella2019energy}.   For each  gate $V_{j}$, the maximum energy increase  is
\begin{align}
e_j:  =  \sup_{|\psi\>}   \,  \<\psi |   V_j^\dag  H_{\rm tot}  V_j  |\psi\>   -  \<\psi |    H_{\rm tot}   |\psi\>   \, \, .
\end{align}
In the following, we will demand that the total energy requirement of the estimation scheme is upper bounded by a constant $E$, independent of $N$.  
Denoting by $e_0$ the initial energy of the state $|\psi\>$,  this requirement leads to the bound  
\begin{align}
  \sum_{j=0}^{2N}  \,   e_j   \le E  \, .   
\end{align}

We now evaluate how the energy increases through the steps of the protocol. Let  $E_n$  be the   energy at the $n$-th step, namely   
\begin{align}\label{En}
E_n   :  =  \<\psi|   \left(  \prod_{j=1}^n    D_{z_j}^\dag V_j^\dag       \right)   \, H_{\rm tot} \, \left(  \prod_{l=1}^n    V_l   D_{z_l}       \right)       |  \psi\>  \,.
\end{align}
The bound (\ref{energyb}) yields the recursion relation  
\begin{align}
\nonumber E_{n+1}    &\le   \left(\sqrt { E_{n} } + \frac {z_{\max}}{\sqrt 2}  \right)^2   +   e_n\\
&  \le \left( \sqrt { E_{n} } + \frac {z_{\max}}{\sqrt 2}  + \sqrt{e_n}\right)^2  \, \,, 
\end{align}
from which we obtain  the bound 
\begin{align}
 \nonumber \sqrt{E_{n}}  &\le      n \,   \frac {  z_{\max}}{\sqrt 2} + \sum_{j=0}^{n}    \sqrt{e_j} \\
 \nonumber &\le   n \,   \frac {  z_{\max}}{\sqrt 2} +   \sqrt {(n +1) \,  \left(  \sum_{j=0}^{n}  e_j\right)  } \\
 \label{energybb}&\le          (2N) \,   \frac {  z_{\max}}{\sqrt 2} +   \sqrt {(2N +1) \, E}  \, .
\end{align}



\subsection{Precision limit for estimation schemes using arbitrary superpositions of definite causal orders}
We now use the  bound  (\ref{energybb}) to prove a precision limit  for  arbitrary schemes using coherent superpositions of causally-ordered setup. 

A generic superposition is specified by a set of causally ordered circuits,  and by the amplitudes assigned to each of these circuits. For the $k$-th term in the superposition,  we denote by  $c_k $ its amplitude, by  $|\psi_k\>$ the initial state of the probe,  by   $(z_{1}^{(k)},  \dots,  z_{2N}^{ (k)})$ the sequence of displacements [equal  to a permutation  of the sequence $(x_1,\dots,  x_N ,  p_1,\dots,  p_N)$], and by $\{  V_j^{(k)}\}_{j=1}^{2N}$ the intermediate gates.   

Denoting by $E^{(k)}$ the energy requirement of the $k$-th quantum circuit, the energy requirement of the superposition is $\sum_k  \,  |c_k|^2  \,  E^{(k)}$.  We demand that the energy requirement is bounded by a constant $E$, independent of $N$, namely  
\begin{align}\label{boundE}
\sum_k  \,  |c_k|^2  \,  E^{(k)} \le E \, .
\end{align}
    
 In the following, we will show that no estimation scheme satisfying Eq. (\ref{boundE}) can have RMSE vanishing faster than $N^{-2}$.

Let us consider the scenario where all the displacements except one, say $x_1$,  are known. In this case,  the quantum Cram\'er-Rao bound reads 
\begin{align}\label{qcrao}
\Delta A  \ge  \frac { 1}{\sqrt{4\nu}    \Delta G}\,,      
\end{align} 
 with $ \Delta  G  :  =  \sqrt{   \<\psi_{\rm in}|   G^2  |\psi_{\rm in} \>   -  \<\psi_{\rm in}|   G  |\psi_{\rm in} \>^2   }$,   $|\psi_{\rm in}\>    :  =  \sum_k  \,  c_k \,   |k\>  \otimes |\psi_k\> $,  $G :  =    i  U^\dag  \frac {\d}{\d A}  U$, and 
 \begin{align}
 U:  =    \sum_k  \,  |k\>\<k|  \otimes \left( \prod_j   \,   V_j^{(k)}  D_{z_j^{(k)}} \right) \,.
 \end{align} 
 Let $j(k)$ be the index such that $z_{j(k)}^{(k)}  =  x_1$.  Then, the gate $U$ can be rewritten as 
 \begin{align}
 U:  =    \sum_k  \,  |k\>\<k|  \otimes   S_k   D_{x_1}    T_k  \, ,    
 \end{align}  
 with $T_k:  =  \prod_{j=1}^{j(k)-1}  V_j^{(k)}    D_{z_j^{(k)}}$ and $S_k:  = \left( \prod_{j=j(k)+1}^{2N}  V_j^{(k)}    D_{z_j^{(k)}} \right) \,  V_{j(k)}^{(k)}$ (adopting the convention $\prod_{j=a}^b A_j:=I$ when $a>b$).  Observing the relation  $  i   D_{x_1}^\dag  \frac {\d}{\d A}    {D_{x_1}}  =    \frac{N}{  \overline p}\,  P $, we then obtain 
 \begin{align}
 G  =  \frac N {   \overline p} \, \sum_k \, |k\>\<k|  \otimes  T_k^\dag       P  T_k  \, ,
 \end{align}
 and therefore
 \begin{align}
\nonumber (\Delta G)^2   &   \le    \<\psi  | G^2  |\psi\>  \\
& =  \frac{  N^2}{\overline p^2}\sum_k   \,  |c_k|^2  \, \<\psi_k  |      T_k^\dag  \, P^2\,   T_k  |\psi_k\>  \,   \,. 
 \end{align}
 Each term in the sum can be upper bounded as 
 \begin{align}
\nonumber  \<\psi_k  |      T_k^\dag  \, P^2\,   T_k  |\psi_k\> &  \le  \<\psi_k  |      T_k^\dag  (X^2+ P^2)  T_k  |\psi_k\> \le    2 E^{(k)}_{j(k)}   \, ,
 \end{align}
 where
 \begin{align}\label{Enk}
E_n^{(k)}   :  =  \<\psi|   \left(  \prod_{j=1}^{j(k)-1}    D_{z^{(k)}_j}^{\dag} V_j^{(k)\dag}       \right)   \, H_{\rm tot} \, \left(  \prod_{l=1}^{j(k)-1}    V_l^{(k)}   D_{z_l^{(k)}}       \right)       |  \psi\>  \,, 
\end{align}
is the total energy at step $j(k)$ in the $k$-th causally-ordered scheme appearing in the superposition.  
 
 Using Equation (\ref{energybb}), we obtain the bound 
 \begin{align}
 \label{boundx} (\Delta G)^2  &\le \frac{ 2N^2 }{  \overline p^2  } \sum_k \,  |  c_k |^2  \,    \left[  2N \,   \frac {  z_{\max}}{\sqrt 2} +   \sqrt { (2N+1)  \,  E^{(k)}  } \right]^2  \, ,
  \end{align}
  where $E^{(k)} $ is the  energy requirement of the $k$-th causally ordered circuit in the superposition. 

By expanding the right-hand-side of (\ref{boundx}), we  then obtain  
\begin{align}
\nonumber (\Delta G)^2  &\le  \frac{2  N^2}{\overline p^2}  \sum_k \,  |  c_k |^2  \,   \left[  4N^2 \,     \frac{z_{\max}^2}2  +    (2N+1)  \,  E^{(k)}     + 4 N \,  \frac{z_{\max} }{\sqrt 2}   \,    \sqrt{  (2N+1)  \,  E^{(k)} }   \right] \\
\nonumber &\le  \frac{2  N^2}{\overline p^2}      \left[  4N^2 \,     \frac{z_{\max}^2}2  +    (2N+1)  \, \<   E\>     + 4 N \,  \frac{z_{\max} }{\sqrt 2}   \,    \sqrt{  (2N+1)  \,  \<E\> }   \right]  \qquad \< E\>   :  =   \sum_k  \,  |c_k|^2  \,  E^{(k)}  
  \\
&  \le  \frac{2  N^2}{\overline p^2}  \,   \left[  2N \,   \frac {  z_{\max}}{\sqrt 2} +   \sqrt { (2N+1)  \,  E    } \right]^2   \, ,
\end{align}
the last equation following from Eq. (\ref{boundE}).

 Inserting the above relation into Equation  (\ref{qcrao}), we finally obtain the precision limit  
\begin{align}
 \Delta  A \ge  \frac {\overline p}{ 4 \sqrt {  \nu}  \,     \,   N^2   \left(  z_{\max}  +   \sqrt { \frac{2N+1}{2N^2}  \, E    }  \right)   \,  } \, ,
\end{align}
showing that every superposition of causally-ordered schemes with bounded energy will have an RMSE vanishing at most as $1/N^2$ in the large $N$ limit.

\end{widetext}

\end{document}